\documentclass[twocolumn]{el-author}

\usepackage[utf8]{inputenc}
\usepackage[bookmarks,colorlinks]{hyperref}
\usepackage{acronym}
\usepackage{times}
\usepackage{amsmath,dsfont}
\usepackage{amssymb,amsthm}

\usepackage{enumerate}
\usepackage{algorithm}  
\usepackage{algpseudocode}

\usepackage{graphicx}

\date{\today}

\usepackage{amsfonts}
\usepackage{dsfont}
\usepackage{color}
\usepackage{graphicx}

\usepackage{tabu}  

\begin{document}

\acrodef{doa}[DOA]{direction-of-arrival}
\acrodef{dl}[DL]{deep learning}
\acrodef{ras}[RAS]{random array selection}
\acrodef{maa}[MAA]{multichannel antenna array}
\acrodef{rmse}[RMSE]{root mean square error}
\acrodef{crb}[CRB]{Cram\'{e}r-Rao lower bound}
\acrodef{cnn}[CNN]{convolution neural network}
\acrodef{music}[MUSIC]{multiple signal classification algorithm}
\acrodef{fista}[FISTA]{fast iterative shrinkage-thresholding algorithm}
\acrodef{rip}[RIP]{restricted isometry property}
\acrodef{snr}[SNR]{signal to noise ratio}
\acrodef{fpga}[FPGA]{field-programmable gate array}
\acrodef{adc}[ADC]{analog to digital converter}
\acrodef{ura}[URA]{uniform rectangular array}
\acrodef{gui}[GUI]{graphical user interface}
\acrodef{mhr}[MHR]{multidimensional harmonic retrieval}	
\acrodef{2d}[2D]{two-dimensional}

\acrodef{cs}[CS]{compressed sensing}	
\acrodef{ista}[ISTA]{iterative shrinkage thresholding algorithm}	
\acrodef{fista}[FISTA]{fast \ac{ista}}

\acrodef{dnn}[DNN]{deep neural network}
\acrodef{rnn}[RNN]{recurrent neural network}

\acrodef{lista}[LISTA]{learned \ac{ista}}	
\acrodef{alista}[ALISTA]{Analytic-LISTA}
\acrodef{adalista}[AdaLISTA]{Adaptive-LISTA}

\acrodef{amp}[AMP]{approximate message passing}

\acrodef{admm}[ADMM]{alternating direction method of multipliers}

\acrodef{snr}[SNR]{signal-to-noise ratio}

\acrodef{nmse}[NMSE]{normalized mean squared error}
\acrodef{2d}[2D]{two-dimensional}

\acrodef{doa}[DOA]{direction-of-arrival}
\acrodef{dft}[DFT]{discrete Fourier transform}
 \vspace{-0.2cm}
\title{Block-Sparse Recovery Network for Two-Dimensional Harmonic Retrieval}
 \vspace{-0.2cm}
\author{Rong Fu, Tianyao Huang, Lei Wang, Yimin Liu}
 \vspace{-0.5cm}
\abstract{
	As a typical signal processing problem, \ac{mhr} has been adapted to a wide range of applications in signal processing.
	Block-sparse signals, whose nonzero entries appearing in clusters, have received much attention recently. 
	An unfolded network, named Ada-BlockLISTA, was proposed to recover a block-sparse signal at a small computational cost, which learns an individual weight matrix for each block. 
	However, as the number of network parameters is increasingly associated with the number of blocks, the demand for parameter reduction becomes very significant, especially for large-scale \ac{mhr}.
	Based on the dictionary characteristics in \ac{2d} harmonic retrieve problems, we introduce a weight coupling structure to shrink Ada-BlockLISTA, which significantly reduces the number of weights without performance degradation. 
	In simulations, our proposed block-sparse reconstruction network, named AdaBLISTA-CP, shows excellent recovery performance and convergence speed in \ac{2d} harmonic retrieval problems.
}

\maketitle

 \vspace{-0.2cm}
\section{Introduction}

Multidimensional harmonic retrieval, associated with many practical applications including \ac{doa} estimation \cite{Balakrishnan2004A,Compbeamf} and range-Doppler estimation \cite{Yuhan2021TSP}, has been extensively studied in the signal processing literature. 
As it is crucial to minimize the required sample size, a myriad of \ac{cs} methods have been developed, such as \ac{ista}.
Moreover, block-sparse signals where the nonzero elements are distributed in clusters arise naturally in many \ac{2d} harmonic retrieval problems.
Thus block-sparse signal recovery has drawn increasing attention and many \ac{cs} methods have been extended in the block-sparse setup by utilizing the block structure, such as Block-ISTA.

	Recently, deep learning methods have gained immense popularity in the field of \ac{cs}. For example, Gregor and LeCun \cite{Gregor2010Learning} have proposed a \ac{rnn} to solve sparse coding problems, named \ac{lista}, which shows improved performance in terms of convergence speed in both theoretical analysis and empirical results \cite{AMP-Inspired,OnsagerLAMP,Fu2021TSP} than traditional iterative solver such as \ac{ista} and \ac{fista}. Due to the success of \ac{lista}, many variants of \ac{lista} have been proposed. For example, robust-ALISTA \cite{liu2018alista} explicitly calculates the learned matrices by solving a coherence minimization problem and only leaves a few network parameters to learn. 
	Another adaptive unfolded network, named \ac{adalista} \cite{2020Ada}, is able to serve different dictionaries using the same weight matrix without retraining.	To further extend the application of \ac{adalista} to block sparse recovery, Ada-BlockLISTA has been proposed--by the same authors of this work--in \cite{2021Deep}, which makes good use of the natural structure of a block-sparse signal thus performs block-wise iterative steps for each separate block.
	As Ada-BlockLISTA learns an individual weight matrix for each block, it enjoys better block recovery performance but suffers from a big increase in the number of network variables, which consumes costly memory/computation resources. Thus, it is important to shrink Ada-BlockLISTA by taking advantage of additional prior knowledge of the signal model.
	Our approach is partly based on the convergence analysis of \cite{liu2018alista,2020Ada} (standard sparse case) and \cite{2021Deep} (block sparse case) where conditions on \textit{learned weights} are provided for the successful recovery in exponential convergence of learned solvers.

In this paper, 
we explore the dictionary structure of \ac{2d} harmonic retrieval and impose weight coupling to the matrices in Ada-BlockLISTA, yielding the design of our AdaBLISTA-CP network.
We also provide numerical results to demonstrate that our coupled network achieves comparable block-sparse recovery performance with Ada-BlockLISTA by using much less network parameters.

 \vspace{-0.2cm}
\section{BRIEF REVIEW OF BLOCK SPARSE RECOVERY}

\textbf{Signal model}:
	The recovery of block sparse signals involves solving a system of linear equations of the form
	\vspace{-0.2cm}
	\begin{equation}  \label{eq:system}
	\bm y =\bm \Phi \bm x^\ast + \bm w,
	\vspace{-0.2cm}
	\end{equation}
	where $ \bm w \in {\mathbb{C}^N}$ is additive random noise in the system and $\bm x^\ast \in {\mathbb{C}^M}, M=PQ$,  is the ground truth, which can be divided into $Q$ sub-vectors as below
	\vspace{-0.2cm}
	\begin{equation}  \label{eq:blockx}
	\bm x = {[
		\underbrace { x_{1,1} \cdots x_{P,1} }_{{\bm x}_1^T}\;
		\underbrace { x_{1,2} \cdots x_{P,2} }_{{\bm x}_2^T}\;\cdots\;
		\underbrace { x_{1,Q} \cdots x_{P,Q} }_{{\bm x}_Q^T}]^T}.
	\vspace{-0.3cm}
	\end{equation}
	The vector $\bm x$ is said to be $K$-block-sparse, if there are at most $K$ nonzero blocks ($K \ll Q$). 
	Sharing the same nested structure with $\bm x$, the dictionary matrix $\bm \Phi$ is also divided into $Q$ blocks, i.e.,
	\vspace{-0.2cm}
	\begin{equation}  \label{eq:blockphi}
	\bm \Phi = [
	\underbrace {\bm \phi _{1,1} \cdots \bm \phi _{P,1}}_{{\bf{\Phi }}_1}\;
	\underbrace {\bm \phi _{1,2} \cdots \bm \phi _{P,2}}_{{\bf{\Phi }}_2}\;\cdots\;
	\underbrace {\bm \phi _{1,Q} \cdots \bm \phi _{P,Q}}_{{\bf{\Phi }}_Q}].
	\vspace{-0.3cm}
	\end{equation}

\textbf{Traditional Iterative Algorithms}:	
	To harness block sparsity, we estimate $\bm x^\ast$ by solving the mixed-norm optimization problem,
	
     \vspace{-0.2cm}
	\begin{equation} \label{eq:blocksparseregression}
	\mathop { \min }\limits_{\bm x} \frac{1}{2}\left\| \bm y - \bm \Phi \bm x \right\|_2^2 + \lambda \left\|\bm x\right\|_{2,1},
	\vspace{-0.2cm}
	\end{equation}
	where $\lambda$ is a regularization parameter controlling the block sparsity penalty characterized by $\ell_{2,1}$ norm defined as $\left\|\bm x\right\|_{2,1} = \sum_{q = 1}^Q \left\|\bm x_q\right\|_{2}$.

	Many algorithms in principle designed for \ac{cs} can be extended to solve block-sparse recovery problems according to the partition of blocks, such as \ac{ista} \cite{Beck2009A}. 
	To solve the $\ell_{2,1}$ minimization problem in \eqref{eq:blocksparseregression}, 
	we briefly review Block-ISTA as an extension of \ac{ista}, which iteratively performs the following two steps (block-wise gradient descent and soft-thresholding) for every block $q \in [1,Q]$:
\vspace{-0.2cm}
\begin{equation}\label{eq:Block-ISTA}
\begin{aligned}
    \bm z_q^{(t+1)}  
    & = {\bm x}_q^{(t)} + \frac{1}{L}{\bm \Phi_q ^H} \left(\bm y  - \bm \Phi {\bm x^{(t)}} \right),\\
    {\bm x}_q^{(t+ 1)} 
    & = {\bm z}_q^{(t+ 1)}{\left( 1 - {\theta} / \left\| {\bm z}_q^{(t+ 1)} \right\|_2 \right)_{ + }},
\end{aligned}
\vspace{-0.2cm}
\end{equation}
where $ (\cdot)_+ $ denotes $ \max(\cdot,0)$, and threshold $ \theta > 0 $ is block-wise soft-thresholding parameter which forces blocks in the updated signal ${\bm z}^{(t+ 1)}$ to $\bm 0$ if its $\ell_{2}$ norm is less than $ {\theta} $.
	Block-ISTA demonstrates considerable accuracy in recovering block sparse signals but takes hundreds or thousands of iterations for convergence.

\textbf{Deep unfolding methods}:		
	Given the outburst in application of \acp{dnn} in \ac{cs}, \ac{lista} and its variants have been proposed to speed up the rate of convergence by freeing the traditional parameters in \ac{ista} to data-driven variables 
	and unfolding \ac{ista} algorithms into a $T$-layer \ac{rnn} ($T \approx 10$). For example, \ac{adalista} is an adaptive version of \ac{lista}, where the single layer computation is \cite{2020Ada}
	\vspace{-0.2cm}
    \begin{equation}
        \label{eq:adalista_singleweight}
        \bm{x}^{(t + 1)} = \eta_{{\theta}^{(t)}}\left( \bm{x}^{(t)} + {\gamma}^{(t)}{\bm \Phi}^H {\bm W}^H \left( \bm y - {\bm \Phi} \bm{x}^{(t)} \right) 
        \right),
    \vspace{-0.2cm}
    \end{equation}
    where ${\theta}^{(t)}$ and ${\gamma}^{(t)}$ are the learned soft threshold and the learned step size at the $t$-th layer. The weight matrix $\bm W$ encodes the structure of $\bm \Phi$, which are shared across different layers. 
    
Recalling the definition of mutual coherence \cite[Definition 3]{2020Ada} and convergence guarantee \cite[Theorem 1,2]{2020Ada}, \ac{adalista} can successfully recover $s$-sparse signals with exponential convergence under the condition that weights satisfies a low enough mutual coherence $\mu(\bm W \bm \Phi, \bm \Phi) = \max_{i \neq j} | \bm \phi^H_{i} {\bm W}^H \bm \phi_j |$, with $\bm \phi^H_{i} {\bm W}^H \bm \phi_i = 1$, where $ \bm \phi_i$ is the $i$-th columns of $\bm \Phi$. 
Furthermore, \cite{liu2018alista} establishes a minimum-coherence criterion between the desired weights and the dictionary, i.e., the optimal learned matrix $\tilde{\bm W}$ is ought to approach the infimum of the generalized mutual coherence. Thus, we have $ {\bm \Phi}^H \tilde{\bm W}^H \bm \Phi \approx \bm I$.

	When it comes to block sparse case, a block-sparse reconstruction network \cite{2021Deep}, named Ada-BlockLISTA, have been proposed.
	Motivated by Block-ISTA, Ada-BlockLISTA makes use of block structure in \ac{adalista} and thus learns an individual weight matrix $ {\bm W}_q$ for each $q$-th block, whose update rule at the $t$-th layer is formulated as
	\vspace{-0.2cm}
	\begin{subequations}\label{eq:AdaBlockLISTA}
    \begin{align}
    {\bm z}_q^{(t+ 1)} 
    & = {\bm x}_q^{(t)} + {\gamma}^{(t)}{\bf{\Phi }}_q^H{\bm W}_q^H \left(\bm y  - \bm \Phi {\bm x^{(t)}} \right), \label{eq:AdaBlockLISTA-sub1}
    \\\vspace{-0.4cm}
    {\bm x}_q^{(t+ 1)} 
    & = {\bm z}_q^{(t+ 1)}{\left( 1 - {\theta ^{(t)}} / \left\| {\bm z}_q^{(t+ 1)} \right\|_2 \right)_{ + }},
    \vspace{-0.2cm}
    \label{eq:AdaBlockLISTA-sub2}
    \vspace{-0.4cm}
    \end{align}
    \end{subequations}
\if false
         \vspace{-0.2cm}
     \begin{equation}\label{eq:AdaBlockLISTA}\
        \begin{aligned}
        {\bm z}_q^{(t+ 1)} 
        & = {\bm x}_q^{(t)} + {\gamma}^{(t)}{\bf{\Phi }}_q^H{\bm W}_q^H \left(\bm y  - \bm \Phi {\bm x^{(t)}} \right), \\
        {\bm x}_q^{(t+ 1)} 
        & = {\bm z}_q^{(t+ 1)}{\left( 1 - {\theta ^{(t)}} / \left\| {\bm z}_q^{(t+ 1)} \right\|_2 \right)_{ + }},
        \end{aligned}
        \vspace{-0.1cm}
        \end{equation}
\fi  
    where $\{{\bm W}_1, \cdots,{\bm W}_Q\},\{\theta^{(t)}, {\gamma}^{(t)}\}_{t=0}^{T}$ are network parameters to learn.
    Based on the definition of sub-coherence and block-coherence \cite{Block-Yonina}, a similar condition of each weight matrix ${\bm W}_q$ in Ada-BlockLISTA ensuring recovery of block-sparse signals can be developed as 
    \vspace{-0.2cm}
    \begin{equation}\label{eq:mini_blockcoherence}
    {\bf{\Phi }}_q^H{\bm W}_q^H \bm \Phi_q \approx \bm I, 
    {\bf{\Phi }}_q^H{\bm W}_q^H \bm \Phi_{q'} \approx \bm 0, q \ne q'.
    \vspace{-0.2cm}
    \end{equation}
    

	Although Ada-BlockLISTA shows both rapid convergence speed and great block-sparse recovery performance, it is difficult to train a large neural network with many weight matrices $\{{\bm W}_q\}_{q=1}^{Q}$, especially when the number of blocks $Q$ is large.
	Therefore, we couple the learned matrices in Ada-BlockLISTA and propose a structured 
	network named AdaBLISTA-CP based on the dictionary characteristics in \ac{2d} harmonic retrieval problems.

\vspace{-0.2cm}
\section{COUPLED NETWORK FOR \ac{2d} HARMONIC RETRIEVAL}
While \ac{adalista} only learns a single weight ${\bm W}$, Ada-BlockLISTA needs to learn a different weight ${\bm W}_q$ for the $q$-th block, whose number of network parameters is up to $\mathcal{O}(Q N^2)$, where $N$ is the number of observation samples. These large matrices consume costly memory/computation resources and require a huge amount of labeled data for training. 
Motivated by the minimum-coherence criterion for block-sparse recovery described in \eqref{eq:mini_blockcoherence}, it is possible to reduce the network parameters by exploring the model structure and introducing some specific relationship between different weights.
In this section, we recall the signal model of \ac{2d} harmonic retrieval problems and reveal its specific characteristics of the dictionary matrices, which helps us couple learned variables in Ada-BlockLISTA.


Following the signal model in \cite{2foldblockToep}, 
\ac{2d} harmonic retrieval problem can be formed into a block-sparse signal estimation problem. We uniformly sample \ac{2d} harmonic frequencies into $P$ and $Q$ points, encapsulated in the set of grid points $ \{p/P\}_{p=0}^{P-1} $ and $ \{q/Q\}_{q=0}^{Q-1} $, respectively.
The full observation matrix $\bm \Psi$ is defined as a \ac{2d} \ac{dft} matrix, given by $
\bm \Psi
= \bm F_{Q} \otimes \bm F_{P},
$
where operator $\otimes$ represents Kronecker product, and $ \bm F_Q $ is a discrete Fourier matrix of size $Q \times Q$, whose $(i, k)$-th entry is $[\bm F_Q]_{i,k}  = \omega_Q^{(i-1)(k-1)}$, where $ \omega_Q = e^{ \mathrm{j} \frac{  2 \pi }{Q} }$, $i,k = 0,1,\cdots,Q-1$. Another discrete Fourier matrix $ \bm F_{P} $ is computed in the same manner.

Considering compressive measurements, we have the dictionary $\bm \Phi \in \mathbb{C}^{N \times M}$ in \eqref{eq:system} consists of $N$ sub-sampled rows of the full dictionary $\bm \Psi$.
To store the indices of the selected rows, we define a subset $\Omega$ of cardinality $N$ randomly chosen from the set $\mathcal{M}:=\{1,2,\dots,M\}$. We use $\bm R$ as a row-sampled matrix to select $N$ rows corresponding to the elements in $\Omega$, i.e., $\left[ \bm R \right]_{n,m} = 1$, where $m$ is the $n$-th element of $\Omega$ while other entries in the $n$-th row are zeros. Thus, the dictionary $\bm \Phi$ can be computed as
$
\bm \Phi
= \bm R  \left( \bm F_{Q} \otimes \bm F_{P} \right).
$

\if false
\begin{equation} \label{eq:Fourier matrix}
\bm F_{M_p} = 
\left[ 
    \begin{array}{*{20}{c}}
		{ 1 } & { 1 } & \cdots & { 1 }\\
		{ 1 } & { \omega }& \cdots & \omega^{M_p-1} \\
		\vdots & \vdots & \ddots & \vdots\\
		{ 1 }& \omega^{M_p-1} & \cdots & \omega^{(M_p-1)(M_p-1)}
	\end{array} 
\right],
\end{equation}
According to the definition of Kronecker product, we have
\begin{equation*}
\bm F_{M_1} \otimes \bm F_{M_2} = 
\left[ 
    \begin{array}{*{20}{c}}
		{\bm C_{1,1}} & {\bm C_{1,2}} & \cdots & {\bm C_{ 1, {M_1} }}\\
		{\bm C_{2,1}} & {\bm C_{2,2}}& \ddots & \vdots \\
		\vdots & \ddots & \ddots &{\bm C_{M_1-1, M_1}}\\
		{\bm C_{{M_1}, 1}}& \cdots &{\bm C_{M_1, M_1-1}}&{\bm C_{M_1, M_1}}
	\end{array} 
\right],
\end{equation*}
where $ \bm C_{i. k} = \omega^{(i-1)(k-1)} \bm F_{M_2}$, $i. k \in \mathcal{M}_1 $, and we have

\begin{equation*}
\left[ 
    \begin{array}{*{20}{c}}
		{\bm I_{M_2}} & {\bm 0} & \cdots & {\bm 0}\\
		{\bm 0} & \!\!\!{\omega^{(k-i)} \bm I_{M_2}}\!\!\! & \ddots & \vdots \\
		\vdots & \ddots & \ddots & {\bm 0}\\
		{\bm 0}& \cdots & \!\!\!\!\!\!{\bm 0}&{\omega^{(M_1-1)(k-i)}\!\!\!\!\!\! \bm I_{M_2}}
	\end{array} 
\right]
\left[ 
    \begin{array}{*{20}{c}}
		{\bm C_{1,i}} \\
		{\bm C_{2,i}} \\
		\vdots \\
		{\bm C_{{M_1}, i}}
	\end{array} 
\right], = 
\left[ 
    \begin{array}{*{20}{c}}
		{\bm C_{1,k}} \\
		{\bm C_{2,k}} \\
		\vdots \\
		{\bm C_{{M_1}, k}}
	\end{array} 
\right].
\end{equation*}

Then we recast \eqref{eq:K_sines_2D} in matrix form as
\begin{eqnarray} \label{eq:MD_dict}
\bm y^{\star} = \bm \Psi \bm x
= \left( \bm F_{M_1} \otimes \bm F_{M_2} \right) \bm x,
\end{eqnarray}
where $\bm x \in \mathbb{C}^M$ contains only $K$ nonzero elements, corresponding to the complex amplitudes of the $K$ sinusoids.


We consider compressive measurements, where only $N$ entries of $\bm y^{\star}$ are observed with $N \ll M$. 
To store the indices of the selected entries from $\bm y^{\star}$, we define a subset $\Omega$ of cardinality $N$ randomly chosen from the set $\mathcal{M}$.
Then we use a row-subsampled matrix $\bm R$ to select $N$ rows of $\bm \Psi$ corresponding to the elements in $\Omega$, i.e., $\left[ \bm R \right]_{n,m} = 1$, where $m$ is the $n$-th element of $\Omega$ while other entries in the $n$-th row are zeros. 
The sub-sampled observations are denoted by $\bm y \in \mathbb{C}^{N}$, with
\begin{equation}
\bm y = \bm R \bm y^{\star}
= \bm R \bm \Psi \bm x,
\end{equation}
Here, we use $ {\bm \Phi} = \bm R \bm \Psi$ to represent the new dictionary matrix

\fi

According to the definition of Kronecker product, we  find that 
each sub-matrix $ {\bf{\Phi }}_q \in {\mathbb{C}^{N \times P}}$ in \eqref{eq:blockphi} can be computed from the first sub-matrix $ {\bf{\Phi }}_1 $ as $ {\bf{\Phi }}_q = {\bm \Lambda} ^{q-1} {\bf{\Phi }}_1$, where ${\bm \Lambda}$ is a diagonal matrix defined as
 \vspace{-0.2cm}
 \begin{tiny}
\begin{equation}
{\bm \Lambda} = \bm R
\left[ 
    \begin{array}{*{20}{c}}
		{\bm I_{P}} & {\bm 0} & \cdots & {\bm 0}\\
		{\bm 0} & {\!\!\! \omega_Q \bm I_{P} \!\!\!}& \ddots & \vdots \\
		\vdots & \ddots & \ddots & {\bm 0}\\
		{\bm 0}& \cdots &{\bm 0}&{\!\!\!\!\!\! \omega_Q^{Q-1} \bm I_{P}\!\!\!}
	\end{array} 
\right]
\bm R.
 \vspace{-0.2cm}
\end{equation}
 \end{tiny}
Thus, \eqref{eq:blockphi} becomes
$
\bm \Phi  = \left[
 \bm \Phi _1, 
 {\bm \Lambda} {\bf{\Phi }}_1, \cdots , 
 {\bm \Lambda} ^{Q-1} {\bf{\Phi }}_1
\right], 
$
where ${\bm \Lambda}^H {\bm \Lambda} = \bm I$.

	
	
\begin{figure}[t]
		\centering
    	\includegraphics[width=0.5\columnwidth]{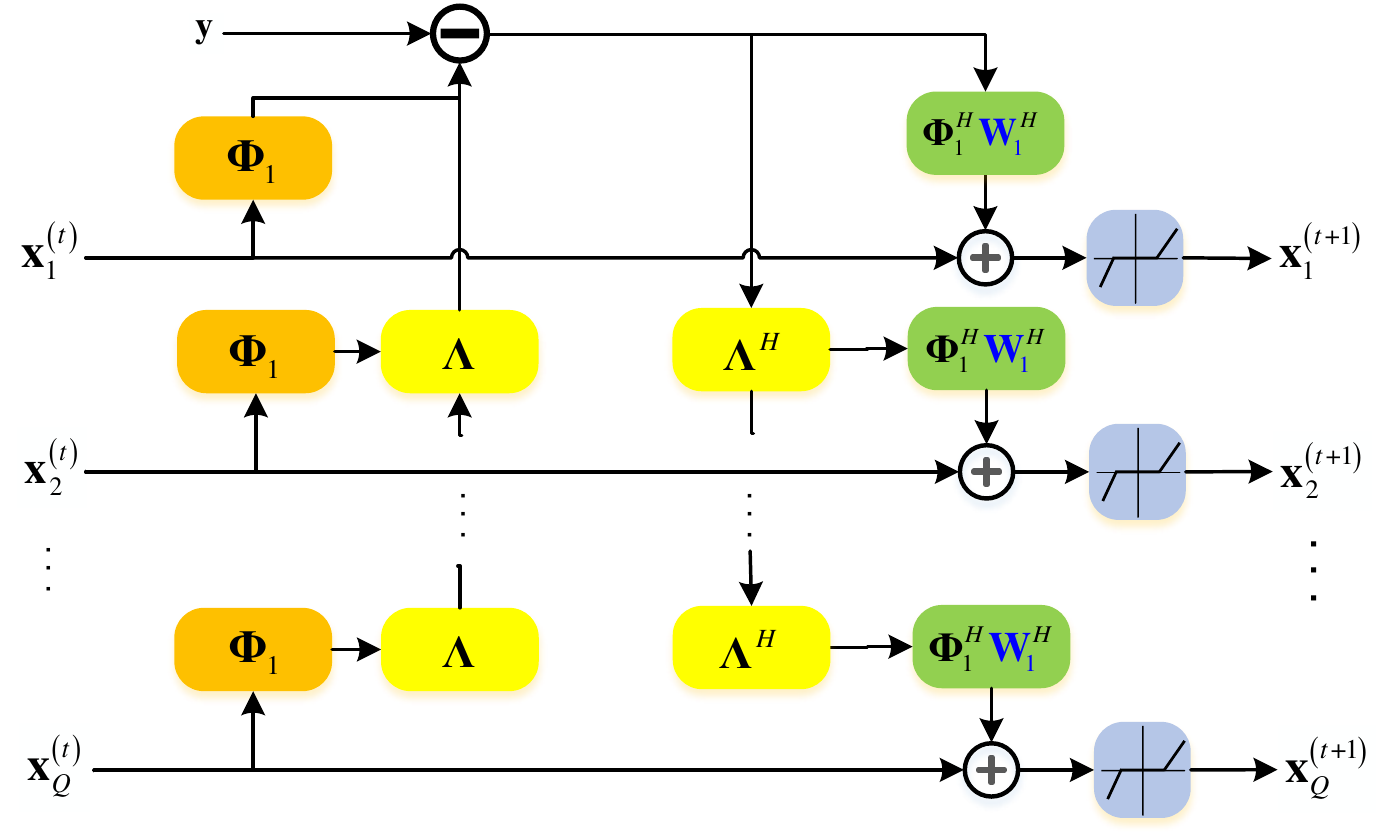}
    	
		\caption{The Block diagram of AdaBLISTA-CP architecture in one layer.}
		\label{fig:frameLISTA}
	\vspace{-0.5cm}
	\end{figure}

	Therefore, as we have ${\bf{\Phi }}_q = {\bm \Lambda} ^{q-1} {\bf{\Phi }}_1$ in \ac{2d} harmonic signal model, we propose a weight coupling method motivated by the minimum-coherence criterion for block-sparse recovery in \eqref{eq:mini_blockcoherence}:
	if $\tilde{\bm W_1}$ satisfies the condition of \eqref{eq:mini_blockcoherence}, then 
	$\tilde{{\bm W}_q} = {\bm \Lambda} ^{q-1} \tilde{\bm W_1} ( {\bm \Lambda}^{q-1})^H$ is also the best parameter for \eqref{eq:mini_blockcoherence}, 
	which motivates us to couple the matrices $\{{\bm W}_q\}_{q=1}^{Q}$ in Ada-BlockLISTA thus leads to a considerable reduction in the number of trained parameters. Therefore, we propose our coupled network, named AdaBLISTA-CP, whose iteration is
	 \vspace{-0.3cm}
 \begin{small}
\begin{equation}\label{eq:AdaBLISTA_CP}
    \begin{aligned}
    {\bm z}_q^{(t+ 1)} 
    & = {\bm x}_q^{(t)} \! + \! {\gamma}^{(t)}{\bf{\Phi }}_1^H {\bm W}_1\!\!^H \!\!( {\bm \Lambda}\!^{q-1} \!)\!^H
    (\bm y \!\! - \!\!  \sum\limits_{i = 1}^Q { {\bm \Lambda} ^{i-1} {\bf{\Phi }}_1 {\bm x}_i^{(t)}} \! ),
    \end{aligned}
     \vspace{-0.3cm}
\end{equation}
 \end{small}
\noindent where we only learn a single weight ${\bm W}_1$ and generate other weights by multiplication with $\bm \Lambda$.
Thus, the number of learned parameters in AdaBLISTA-CP is reduced to $\mathcal{O}(N^2)$, which contributes to lower storage burden and higher sample efficiency.
We illustrate the network structure of AdaBLISTA-CP in Fig.~\ref{fig:frameLISTA}.

\if false
\begin{figure} [tb]
	\centering
	\includegraphics[width=0.3\columnwidth]{figure/blockGramMat.pdf}
	\caption{Visualization of the Gram matrix ${\bm \Phi}^H{\bm \Phi}$ with block-structure.}
	\label{fig:gram}
\end{figure}

define a set of ``good'' weights which $W^k$s are chosen from:
\begin{definition}
Given $A\in \Re^{m \times n}$, a weight matrix is ``good'' if it belongs to
\begin{equation}
\label{eq:optimal_w}
W(A) = \argmin_{W \in \Re^{m \times n}}
\bigg\{ \max_{1 \leq i,j \leq n }|W_{i,j}|: (W_i)^TA_i = 1,1 \leq i \leq n, \max_{\substack{i\neq j\\1 \leq i,j \leq n} }|(W_i)^\top A_j| = \tilde{\mu} \bigg\}.
\end{equation}
Let $C_W = \max_{1 \leq i,j \leq n }|W_{i,j}|$, if $W \in \X_W(A)$.
\end{definition}
With definitions (\ref{eq:mutual_coher}) and (\ref{eq:optimal_w}), we propose a choice of parameters:
\begin{equation}
\label{eq:wtheta}
W^k \in \X_W(A), \quad \theta^k =  \sup_{(x^*,\varepsilon)\in \X(B,s,\sigma)}\{\tilde{\mu} \|x^k(x^*,\varepsilon) - x^*\|_1\} + C_W \sigma,
\end{equation}
which are uniform for all $(x^*,\varepsilon)\in \X(B,s,\sigma)$. In the following proof line, we prove that (\ref{eq:wtheta}) leads to the conclusion (\ref{eq:linear_conv}) in Theorem \ref{prop:no_ss}.

By adopting the above weight coupling, we reform \eqref{eq:AdaBlockLISTA-sub1} of Ada-BlockLISTA and construct our AdaBLISTA-CP:
\begin{equation}\label{eq:AdaBlockLISTA_CP}
    \begin{aligned}
    {\bm z}_q^{(t+ 1)} 
    & = {\bm x}_q^{(t)} + {\gamma}^{(t)}{\bf{\Phi }}_q^H {\bm \Lambda} ^{q-1} {\bm W}_1^H ( {\bm \Lambda}^{q-1})^H \left(\bm y - \sum\limits_{i = 1}^Q { {{\bm \Phi}_i}{\bm x}_i^{(t)}} \right) \\
    & = {\bm x}_q^{(t)} + {\gamma}^{(t)}{\bf{\Phi }}_1^H {\bm W}_1^H ( {\bm \Lambda}^{q-1})^H \left(\bm y - \sum\limits_{i = 1}^Q { {\bm \Lambda} ^{i-1} {\bf{\Phi }}_1 {\bm x}_i^{(t)}} \right),
    \end{aligned}
\end{equation}

\begin{table} 
	
	\begin{tabu} to 0.5\textwidth{X[4,c]|X[4,c]|X[5,c] |X[6,c]} 
		\hline 
		Block-ISTA & \ac{adalista} & AdaBLISTA & \textbf{AdaBLISTA-CP}\\ 
		\hline 
		$\mathcal{O}(1/{\epsilon})$ &
		\multicolumn{3}{c} {
		$\mathcal{O}(\log(1/{\epsilon}))$ } \\ 
		\hline 
		- &
		$O(N^2+T)$ &
		$O(Q N^2+T)$ &
		$O(N^2+T)$ \\ 
		\hline 
	\end{tabu} 
	\caption{The convergence rate (first row) and  and the number of parameters to learn (second row).} 
	\label{table1}
\end{table}

\begin{table} 
		\caption{The convergence rate 
		and the number of parameters to learn
		} \label{table1}
		\begin{tabu} to 0.5\textwidth{X[6,c]|X[5,b]|X[5,l]} 
			\hline 
			Network & convergence rate & the number of parameters to learn \\ 
			\hline 
			Block-ISTA & $\mathcal{O}(1/{\epsilon})$ & -\\
			\ac{adalista} & $\mathcal{O}(\log(1/{\epsilon}))$ & $\mathcal{O}(N^2+T)$ \\
			Ada-BlockLISTA & $\mathcal{O}(\log(1/{\epsilon}))$ & $\mathcal{O}(Q N^2+T)$ \\
			\textbf{AdaBLISTA-CP} & $\mathcal{O}(\log(1/{\epsilon}))$ & $\mathcal{O}(N^2+T)$ \\
			\hline 
		\end{tabu} 
\end{table} 
\fi	
	
\vspace{-0.2cm}
\section{NUMERICAL RESULTS}
We compare the performance of three deep unfolding networks (\ac{adalista}, Ada-BlockLISTA, and our AdaBLISTA-CP) in block sparse recovery. 
In our simulation, we generate noisy observed signals according to \eqref{eq:system} 
where the block-sparse signal $\bm x ^\ast \in {\mathbb{C}^{P Q}} $ has $Q = 64$ blocks each with block size $P = 4$, and the number of non-zero blocks in $\bm x$ is $K \in \{ 1, 2, 3, 4, 5 \}$.
	Note that the inputs to our reconstruction network are complex-value data, thus we transform every operator above to its complex value counterparts by following complex-value extension methods presented in \cite{Fu2021TSP}. 

\begin{figure} [tb]
		\centering
		\includegraphics[width=0.9\columnwidth]{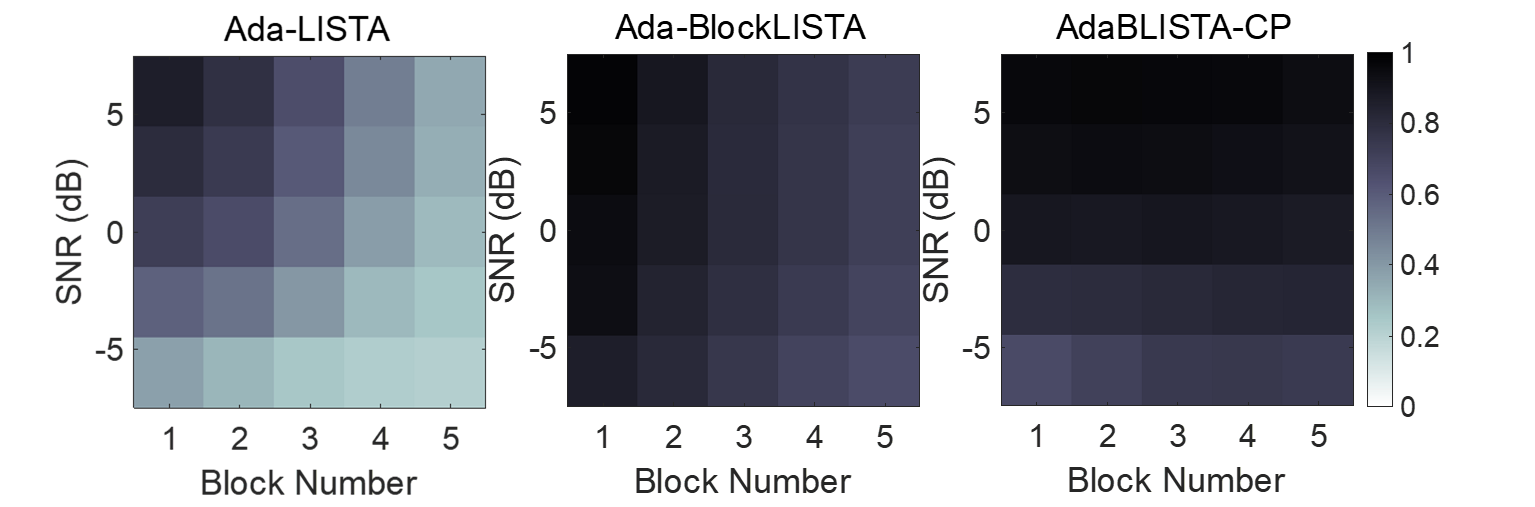}
		\vspace{-0.31cm}
		\caption{Hit rate of three networks with various block sparsity and noise power.}
		\label{fig:hit rate}
	\vspace{-0.5cm}
	\end{figure}

As shown in Fig.~\ref{fig:hit rate}, we evaluate the support recovery performance of block-sparse signals in terms of hit rate versus \ac{snr} and block sparsity.
	The \ac{snr} is computed as $\mathrm{SNR} = 10\log _{10} \frac{1}{
		\sigma^2 }$, where $\sigma^2$ is the noise variance.
	The hit rate is defined as the percentage of successes in finding the nonzero blocks in $\bm x$ against noise.
In Fig.~\ref{fig:hit rate}, a larger area of the dark color part represents better block-sparse recovery performance, which demonstrates that block-sparse reconstruction networks (Ada-BlockLISTA and AdaBLISTA-CP) have better block-sparse recovery performance than the non-block counterpart (Ada-LISTA) when it comes to high noise power and a large number of blocks.
Furthermore, because our proposed network takes advantage of dictionary structure to couple network parameters, it is much easier to learn its best network parameters with limited labeled data and relatively low hardware expenses. 
	Therefore, AdaBLISTA-CP enjoys strong robustness to the block sparsity and measurement noise as well as linear convergence rate with less network parameters.
	
	%
	
\vspace{-0.2cm}
\section{CONCLUSION}
In this work, we considered the block-sparse signal model in \ac{2d} harmonic retrieval problem, where nonzero entries of recovered signal occur in clusters, and derived our AdaBLISTA-CP network by leveraging the particular dictionary structure in the signal model. 
AdaBLISTA-CP inherits the structure of dictionary and reduces the number of parameters to train by exploring the relationship between different weights across blocks. 
In simulation, our proposed network with coupled weight matrices shows excellent recovery performance in terms of hit rate, better than Ada-LISTA and comparable to Ada-BlockLISTA.

\vspace{-0.3cm}
\vskip3pt
\ack{This project is funded by the National Natural Science Foundation of China under Grants No. 61801258 and 62171259.}

\vskip5pt
\vspace{-0.2cm}
\noindent Rong Fu, Tianyao Huang, Lei Wang, Yimin Liu
(\textit{
	Department of Electrical Engineering, Tsinghua University, Beijing, China})
\vskip3pt

\noindent Correspondence: huangtianyao@tsinghua.edu.cn

\vspace{-0.2cm}
\bibliographystyle{IEEEtran}
\bibliography{IEEEabrv}

\begin{thebibliography}{10}
\providecommand{\url}[1]{#1}
\csname url@samestyle\endcsname
\providecommand{\newblock}{\relax}
\providecommand{\bibinfo}[2]{#2}
\providecommand{\BIBentrySTDinterwordspacing}{\spaceskip=0pt\relax}
\providecommand{\BIBentryALTinterwordstretchfactor}{4}
\providecommand{\BIBentryALTinterwordspacing}{\spaceskip=\fontdimen2\font plus
\BIBentryALTinterwordstretchfactor\fontdimen3\font minus
  \fontdimen4\font\relax}
\providecommand{\BIBforeignlanguage}[2]{{%
\expandafter\ifx\csname l@#1\endcsname\relax
\typeout{** WARNING: IEEEtran.bst: No hyphenation pattern has been}%
\typeout{** loaded for the language `#1'. Using the pattern for}%
\typeout{** the default language instead.}%
\else
\language=\csname l@#1\endcsname
\fi
#2}}
\providecommand{\BIBdecl}{\relax}
\BIBdecl

\bibitem{Balakrishnan2004A}
R.~D. Balakrishnan and H.~M. Kwon, ``A new inverse problem based approach for
  azimuthal doa estimation,'' in \emph{Proc. IEEE Global Commun. Conf. (IEEE
  GLOBECOM)}, vol.~4, 2004, pp. 2187--2191.

\bibitem{Compbeamf}
A.~Xenaki, P.~Gerstoft, and K.~Mosegaard, ``Compressive beamforming,''
  \emph{The Journal of the Acoustical Society of America}, vol. 136, p. 260, 07
  2014.

\bibitem{Yuhan2021TSP}
Y.~Li, T.~Huang, X.~Xu, Y.~Liu, L.~Wang, and Y.~C. Eldar, ``Phase transitions
  in frequency agile radar using compressed sensing,'' \emph{IEEE Transactions
  on Signal Processing}, vol.~69, pp. 4801--4818, 2021.

\bibitem{Gregor2010Learning}
K.~Gregor and Y.~Lecun, ``Learning fast approximations of sparse coding,'' in
  \emph{International Conference on International Conference on Machine
  Learning}, 2010, pp. 399--406.

\bibitem{AMP-Inspired}
M.~{Borgerding}, P.~{Schniter}, and S.~{Rangan}, ``{AMP}-inspired deep networks
  for sparse linear inverse problems,'' \emph{IEEE Transactions on Signal
  Processing}, vol.~65, no.~16, pp. 4293--4308, Aug 2017.

\bibitem{OnsagerLAMP}
M.~Borgerding and P.~Schniter, ``Onsager-corrected deep learning for sparse
  linear inverse problems,'' in \emph{2016 IEEE Global Conference on Signal and
  Information Processing (GlobalSIP)}, Dec 2016, pp. 227--231.

\bibitem{Fu2021TSP}
R.~Fu, Y.~Liu, T.~Huang, and Y.~C. Eldar, ``Structured lista for
  multidimensional harmonic retrieval,'' \emph{IEEE Transactions on Signal
  Processing}, vol.~69, pp. 3459--3472, 2021.

\bibitem{liu2018alista}
\BIBentryALTinterwordspacing
J.~Liu, X.~Chen, Z.~Wang, and W.~Yin, ``{ALISTA}: Analytic weights are as good
  as learned weights in {LISTA},'' in \emph{International Conference on
  Learning Representations}, 2019. [Online]. Available:
  \url{https://openreview.net/forum?id=B1lnzn0ctQ}
\BIBentrySTDinterwordspacing

\bibitem{2020Ada}
A.~Aberdam, A.~Golts, and M.~Elad, ``Ada-lista: Learned solvers adaptive to
  varying models,'' \emph{arXiv:2001.08456}, 2020.

\bibitem{2021Deep}
R.~Fu, V.~Monardo, T.~Huang, and Y.~Liu, ``Deep unfolding network for
  block-sparse signal recovery,'' in \emph{ICASSP 2021 - 2021 IEEE
  International Conference on Acoustics, Speech and Signal Processing
  (ICASSP)}, 2021, pp. 2880--2884.

\bibitem{Beck2009A}
A.~Beck and M.~Teboulle, ``A fast iterative shrinkage-thresholding algorithm
  for linear inverse problems,'' \emph{Siam J Imaging Sciences}, vol.~2, no.~1,
  pp. 183--202, 2009.

\bibitem{Block-Yonina}
Y.~C. Eldar, P.~Kuppinger, and H.~Bolcskei, ``Block-sparse signals: Uncertainty
  relations and efficient recovery,'' \emph{IEEE Transactions on Signal
  Processing}, vol.~58, no.~6, pp. 3042--3054, 2010.

\bibitem{2foldblockToep}
Y.~{Chi} and Y.~{Chen}, ``Compressive two-dimensional harmonic retrieval via
  atomic norm minimization,'' \emph{IEEE Transactions on Signal Processing},
  vol.~63, no.~4, pp. 1030--1042, Feb 2015.

\end{thebibliography}

\end{document}